\documentclass[aps,prl,twocolumn,groupedaddress]{revtex4-1}
\usepackage[pdftex]{graphicx}
\usepackage{amsmath}    
\usepackage{subfigure}  
\usepackage[english]{babel}
\usepackage{braket}
\usepackage{epstopdf}

\begin{document}


\title{Super-Coulombic atom-atom interactions \\  in hyperbolic media}

\author{Cristian L. Cortes$^{1,2}$, Zubin Jacob$^{1,2}$}
\email[]{zjacob@purdue.edu}
\affiliation{Department of Electrical and Computer Engineering, \\ University of Alberta, Edmonton, AB T6G 2V4, Canada}
\affiliation{Birck Nanotechnology Center and Purdue Quantum Center, \\ School of Electrical and Computer Engineering,\\ Purdue University, West Lafayette, IN 47906, U.S.A.}
\hspace{8cm}

\begin{abstract}

Dipole-dipole interactions which govern phenomena like cooperative Lamb shifts, superradiant decay rates, Van der Waals forces, as well as resonance energy transfer rates are conventionally limited to the Coulombic near-field. Here, we reveal a class of real-photon and virtual-photon long-range quantum electrodynamic (QED) interactions that have a singularity in  media with hyperbolic dispersion. The singularity in the dipole-dipole coupling, referred to as a Super-Coulombic interaction, is a result of an effective interaction distance that goes to zero in the ideal limit irrespective of the physical distance. We investigate the entire landscape of atom-atom interactions in hyperbolic media and propose practical implementations with phonon-polaritonic hexagonal boron nitride in the infrared spectral range and plasmonic super-lattice structures in the visible range. Our work paves the way for the control of cold atoms in hyperbolic media and the study of many-body atomic states where optical phonons mediate quantum interactions. 
\end{abstract}


\maketitle

Dipole-dipole interactions (DDI) are instrumental in mediating entanglement and super-radiance in cold atoms  \cite{bloch2008quantum,wang2007superradiance}, coherent coupling between single molecules or single atoms \cite{hettich2002nanometer,goban2015superradiance,gonzalez2015subwavelength,ravets2014coherent}, virtual photon exchange and frequency shifts in circuit QED \cite{van2013photon,majer2007coupling}, and F\"orster resonance energy transfer transfer between dye molecules or quantum dots \cite{andrew2004energy,blum_nanophotonic_2012}. There are two fundamental ways of controlling the strength and length scales of dipole-dipole interactions. The first method involves the tuning of intrinsic atomic properties such as transition dipole moments and transition frequencies with highly-excited Rydberg atoms and superconducting qubits \cite{ravets2014coherent,saffman2010quantum,devoret2004superconducting}. The second method involves the tuning of the quantum electrodynamic vacuum, achieved through cavities, waveguides and photonic bandgaps  \cite{fleury2013enhanced,maas2013experimental,martin2010resonance,biehs2013large}. This opens the important question whether a homogeneous material or metamaterial can also enhance dipole-dipole interactions. \\
In this Letter, we reveal a class of divergent excited-state atom-atom interactions that can occur in natural and artificial media with hyperbolic dispersion. Unlike the above mentioned approaches which engineer radiative coupling, we show that the homogeneous hyperbolic medium itself fundamentally alters the Coulombic near-field. The resultant singular long-range interaction, referred to as a Super-Coulombic interaction, is described by an effective interaction distance that goes to zero ($r_e\rightarrow 0$) along a material-dependent resonance angle. We show that this interaction affects the entire landscape of real photon and virtual photon phenomena such as the cooperative Lamb shift, the cooperative decay rate, resonance energy transfer rates and frequency shifts as well as resonant interatomic forces. While we find that the singularity is curtailed by material absorption, it still allows for interactions with much larger magnitudes and longer ranges than those found in any conventional media. We also show that atoms in a hyperbolic medium will exhibit a strong orientational dependence that can effectively switch the dipolar interaction off or on,  providing an additional degree of freedom to control DDI. Our investigation reveals a marked contrast between ground-state and excited-state interactions which can be used to distinguish the Super-Coulombic effect in experiment. Finally, we propose practical broadband hyperbolic systems consisting of phonon-polaritons and plasmonic-media to experimentally verify our predicted Super-coulombic effect. 


\begin{figure}[t!]
\begin{center}
\includegraphics[width=70mm]{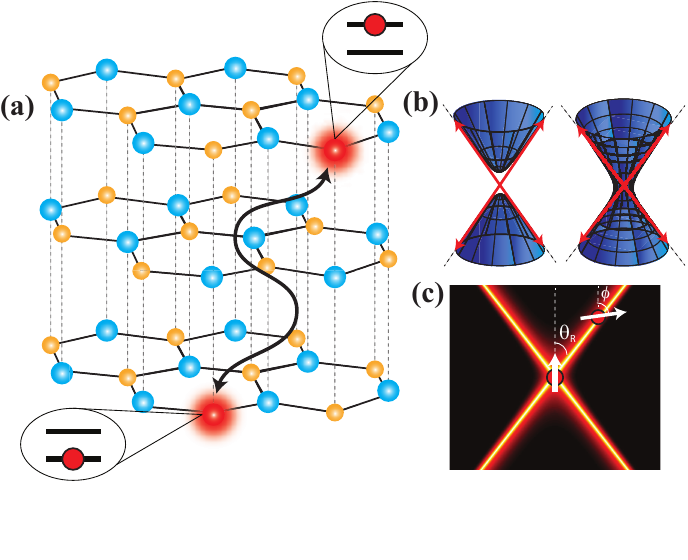}
\end{center}
\vspace{-40px}

\caption{(a) Two dopant atoms in a hyperbolic medium (e.g. h-BN) will interact via a long range super-Coulombic dipole-dipole interaction (DDI). (b)-(c) The DDI occurs over a broad range of frequencies along the resonance cone angle of a hyperbolic medium and causes the effective interaction distance to approach zero irrespective of the physical distance.
}
\label{figure1} 
\vspace{-10px}
\end{figure}

We emphasize that the materials platform we introduce in this paper to enhance dipole-dipole interactions is fundamentally different from the cavity QED \cite{imamog1999quantum,ye1999trapping} or waveguide QED regimes \cite{zheng2010waveguide,van2013photon,goban2015superradiance,chang2007single}. We do not rely on atom confinement \cite{ye1999trapping,goban2015superradiance,gonzalez2015subwavelength,ravets2014coherent,wang2007superradiance}, cavity resonances or modal effects such as the quasi-TEM mode in circuit QED  \cite{van2013photon}, the band-edge slow light as in PhC waveguides \cite{baba2008slow,goban2015superradiance,gonzalez2015subwavelength}, low mode volume plasmonic waveguides \cite{chang2007single,gonzalez2010entanglement} or infinite phase velocity at a cut-off frequency as in ENZ waveguides \cite{fleury2013enhanced,sokhoyan2013quantum}. We also stress that the Super-Coulombic effect engineers the conventional non-radiative fields as opposed to radiative modes and will occur over a broad range of frequencies due to the broadband nature of the hyperbolic dispersion relation \cite{poddubny2013hyperbolic,cortes2012quantum}. Figure 1 depicts a schematic of the proposed Super-Coulombic dipole-dipole interaction using hexagonal Boron Nitride (h-BN) \cite{caldwell2014sub} and two dopant atoms. In the infrared spectral range, h-BN is a uniaxial material that supports ordinary waves (polarization perpendicular to the optic axis) and extraordinary waves (polarization along the optic axis). Extraordinary waves satisfy the hyperbolic dispersion relation $k_x^2/\epsilon_z + k_z^2/\epsilon_x = \omega^2/c^2$ when $\epsilon_x\epsilon_z < 0 $.\\
We begin by formulating the QED theory \cite{knoll2000qed} of dipole-dipole interactions between two neutral, non-magnetic atoms in a hyperbolic medium. We focus on dipolar interactions where the electrodynamic field is initially prepared in the vacuum state $\ket{\{0\}}$. Using the multipolar Hamiltonian, the interaction of two neutral atoms [positions $\mathbf{r}_j$, transition frequencies $\omega_j$ and transition electric dipole moments $\hat{\mathbf{d}}_j$ ($j=A,B$)] is specified by the interaction Hamiltonian $\hat H_{\text{int}}=-\sum_j\int_0^\infty \!d\omega[\mathbf{\hat d}_j\mathbf{\hat E}(\mathbf{r}_j,\omega) + h.c.]$, where $h.c.$ stands for the Hermitian conjugate. The matter-assisted electric field is given by $\mathbf{\hat E}(\mathbf{r},\omega) = i\sqrt{\hbar\omega/\pi\epsilon_oc} \int d^3\mathbf{r}' \sqrt{\epsilon''(\mathbf{r}',\omega)} \mathbf{G}(\mathbf{r},\mathbf{r}',\omega)\mathbf{\hat f}(\mathbf{r}';\omega)$
where $\mathbf{G}(\mathbf{r},\mathbf{r}';\omega)$ is the classical dyadic Green function that satisfies the macroscopic Maxwell equations. Here, $\mathbf{\hat f}^\dagger(\mathbf{r}',\omega)$ and $\mathbf{\hat f}(\mathbf{r}',\omega)$ are bosonic field operators which play the role of the creation and annihilation operators of the matter-assisted electromagnetic (polaritonic) field. The unique interaction properties are a direct result of the dispersion relation of the hyperbolic polariton, as opposed to the photonic dispersion relation, $\omega=ck$, seen in vacuum. The electric field is defined so that it rigorously satisfies the equal-time commutation relations and fluctuation-dissipation theorem \cite{knoll2000qed}. We use conventional perturbation theory to calculate the various dipolar interactions in a hyperbolic medium. We emphasize that the QED theory captures both ground state-ground state interactions and excited state-ground state interactions which a semiclassical approach cannot. \\
If the initial state of the atomic system is prepared in the symmetric or anti-symmetric state, $\ket{i}=\frac{1}{\sqrt{2}}(\ket{e_A}\ket{g_B}\pm\ket{g_A}\ket{e_B})$, then one can show that the resonant dipole-dipole interaction (RDDI) [see Supp. Info.] is given by
\begin{align}
	V_{dd} &= \hbar\left(J_{dd} - i\frac{\gamma_{dd}}{2}\right) = -\frac{\omega_A^2}{\epsilon_o c^2} \mathbf{d}_B\cdot\mathbf{G}(\mathbf{r}_B,\mathbf{r}_A;\omega_A)\cdot\mathbf{d}_A.
	\label{RDDI}
\end{align}
where $\mathbf{d}_j=\braket{g_j|\hat{\mathbf{d}}_j|e_j}$ is the transition dipole moment of atom $j$, assumed to be real. $J_{dd}$ is the cooperative Lamb shift (also known as the virtual photon exchange interaction) and $\gamma_{dd}$ is the cooperative decay rate commonly associated with superradiant or subradiant effects. 


Our result for the resonant dipole-dipole interaction in a hyperbolic medium ($\boldsymbol{\epsilon}=\text{diag}[\epsilon_x,\epsilon_x,\epsilon_z]$) is
\begin{equation}
	V_{dd} =\! \frac{e^{ik_o r_e}}{4\pi\epsilon_o\sqrt{\epsilon_{x}}r_e^3}\mathbf{d}_B \cdot\!\left[(1-ik_or_e)\boldsymbol{\kappa}_{nf}\!- k_o^2r_e^2\boldsymbol{\kappa}_{ff} \right]\!\cdot\mathbf{d}_A   + \tilde{V}_{dd}^{eo}
	\label{GF_aniso}
\end{equation}
valid when $\mathbf{r}_A\neq\mathbf{r}_B$. The first term arises exclusively from extraordinary waves following a hyperbolic dispersion while the second term $\tilde{V}_{dd}^{eo}$ arises from a combination of ordinary and extraordinary waves [SI]. Here, we have defined the near-field and far-field dipole orientation matrix factors $\boldsymbol{\kappa}_{nf}=\epsilon_{x}\epsilon_{z}(\boldsymbol{\epsilon}^{-1} - 3 (\boldsymbol{\epsilon}^{-1}\cdot\mathbf{r})(\boldsymbol{\epsilon}^{-1}\cdot\mathbf{r})/(\mathbf{r}\cdot\boldsymbol{\epsilon}^{-1}\cdot\mathbf{r}))$ and $\boldsymbol{\kappa}_{ff} = \epsilon_{x}\epsilon_{z}(\boldsymbol{\epsilon}^{-1} - (\boldsymbol{\epsilon}^{-1}\cdot\mathbf{r})(\boldsymbol{\epsilon}^{-1}\cdot\mathbf{r})/(\mathbf{r}\cdot\boldsymbol{\epsilon}^{-1}\cdot\mathbf{r}))$ respectively. Equation (\ref{GF_aniso}) reduces to the vacuum RDDI expression when $\epsilon_{x}=\epsilon_{z}=1$, which is applicable both in the retarded ($r\gg\lambda$) and non-retarded ($r\ll\lambda$) regimes. The most unique aspect of dipole-dipole interactions in uniaxial media is the divergence that is predicted from the first term only when the hyperbolic condition ($\epsilon_{x}\epsilon_{z}<0$) is satisfied. In the ideal lossless limit, we find that the effective interaction distance between two atoms, $r_e = \sqrt{\epsilon_x\epsilon_z(\mathbf{r}\cdot\boldsymbol\epsilon^{-1}\cdot\mathbf{r})} = r \sqrt{\epsilon_{z}\sin^2\theta+\epsilon_{x}\cos^2\theta}$, tends towards the limit
\begin{equation}
	r_e \rightarrow 0 \;\; \text{as} \;\; \theta\rightarrow \theta_R=\tan^{-1}\sqrt{-\epsilon_{x}/\epsilon_{z}}. 
\end{equation}
This super--coulombic effect results in the divergence of the dipole-dipole interaction strength $|V_{dd}|/\hbar$ along the  resonance angle $\theta_R$, defined with respect to the optic axis. 

Atoms in a hyperbolic medium will then have an associated cooperative Lamb shift (CLS) and cooperative decay rate  (CDR)
\begin{align}
	J_{dd} &\approx \frac{\sqrt{\epsilon_x}\epsilon_{z}}{4\pi\hbar\epsilon_o r_e^3}\,\mathbf{d}_B\cdot\![\boldsymbol{\epsilon}^{-1}\! - 3\frac{(\boldsymbol{\epsilon}^{-1}\!\cdot\mathbf{r})(\boldsymbol{\epsilon}^{-1}\!\cdot\mathbf{r})}{\mathbf{r}\cdot\boldsymbol{\epsilon}^{-1}\cdot\mathbf{r}}]\!\cdot\mathbf{d}_A
	\label{Lambshift} \\
	\gamma_{dd} &\approx \frac{\omega_A^3\sqrt{\epsilon_x}\epsilon_{z} }{3\pi\hbar\epsilon_oc^3}\mathbf{d}_B\cdot\!\boldsymbol{\epsilon}^{-1}\cdot\!\mathbf{d}_A
	\label{DecayRate}
\end{align}
%
%
in the limit $\theta\rightarrow\theta_R$. Equations (4) and (5) are the dominant factors of the extraordinary wave contribution only. 

This scaling behavior mediated by hyperbolic modes is in striking contrast to the behavior of  the cooperative Lamb shift in vacuum. In vacuum, the CLS depends crucially on the interatomic distance. For separation distances much larger than the transition wavelength, the CLS scales as $J_{dd}\sim \gamma_o \cos(k_o r)/(k_o r)$ and becomes much smaller than the free-space spontaneous emission rate $\gamma_o$. For distances much smaller than the wavelength, the CLS scales as $J_{dd} \sim \gamma_o/(k_o r)^3$, which implies that it can become much larger than the spontaneous emission rate. In contrast, the cooperative Lamb shift in a hyperbolic medium is dependent on $r_e^{-3}$, $J_{dd}\sim \gamma_o/(k_o r_e)^3$, for all interatomic distances. The material-dependent factor 1/$r_e^3$ diverges in the lossless case and therefore results in  giant dipole-dipole interactions for short and large interatomic distances.  

This contrast is also revealed in the cooperative decay rate. At large distances, the CDR in vacuum scales as $\gamma_{dd}\sim \gamma_o \sin(k_o r)/(k_o r)$, therefore becoming weak for distances much larger than the wavelength. For distances much smaller than the wavelength, the CDR becomes independent of position, $\gamma_{dd} \sim \gamma_o$, and remains on the order of the free space spontaneous emission rate. In contrast, the cooperative decay rate in a hyperbolic medium along the resonance angle is not dependent on the effective interaction distance $r_e$, and instead it depends crucially on the orientation angle $\phi$ of the dipoles, $\gamma_{dd} \sim \gamma_o (\epsilon_{z}/\sqrt\epsilon_x \sin^2\phi +  \sqrt\epsilon_x\cos^2\phi )$. When both dipoles are oriented perpendicular to the optic axis ($\phi=\pi/2$), there exists a unique wavelength when the medium can achieve an anisotropic epsilon-near-zero (ENZ) medium ($\epsilon_x\rightarrow 0$ and $\epsilon_z\neq 0$) resulting in a divergent cooperative decay rate, independent of interatomic distance. When both dipoles are parallel to the optic axis ($\phi=0$), the same anisotropic ENZ condition gives a null CDR between the two atoms, independent of interatomic distance. 

We will now consider the role of material absorption ($\epsilon_{x}=\epsilon_{x}'+i\epsilon_{x}''$ and $\epsilon_{z}=\epsilon_{z}'+i\epsilon_{z}''$) on atom-atom interactions in a hyperbolic medium. We find that the effective interaction distance tends to the finite value
 $|r_e| \rightarrow |r|\left[\frac{\epsilon_{z}''|\epsilon_{x}'|+\epsilon_{x}''|\epsilon_{z}'|}{|\epsilon_{x}'|+|\epsilon_{z}'|} \right]^{1/2} \text{as} \;\; \theta\rightarrow \theta_R, $ which curtails the singularity of the hyperbolic dipolar interaction but nevertheless allows for very large interaction strengths compared to conventional media whenever $|r_e|/|r|<1$ is satisfied. The second important consequence of material absorption results from the mixing of terms in the real and imaginary components of the RDDI in eqn. (\ref{GF_aniso}).  Both the cooperative Lamb shift and decay rate scale as $r_e^{-3}$ thus becoming large-yet-finite values in the small interatomic distance limit. The third consequence of material absorption on resonant dipole-dipole interactions is in the transition from non-retarded ($r^{-3}$) to retarded ($r^{-1}$) interactions. In vacuum, the transition occurs when the interatomic separation distance is on the order of wavelength,$(\omega/c)r \sim 1$. Along the resonance angle of a hyperbolic medium, the transition is expected to occur approximately when $(\omega/c)|r_e|\sim 1$. It should be noted that in the lossless limit, this condition is never satisfied since $r_e\rightarrow 0$ for ideal hyperbolic media. Therefore we find that RDDI should scale with the characteristic power law of near-field non-radiative interactions ($r^{-3}$) for all interatomic distances. Once material absorption is included, dipolar interactions will transition from the power law ($r_e^{-3}$ ) to the exponential scaling law ($e^{-(\omega/c)\text{Im}[r_e]}$) valid at large interatomic distances.
\begin{figure}[t!]
\begin{center}
\includegraphics[width=86mm]{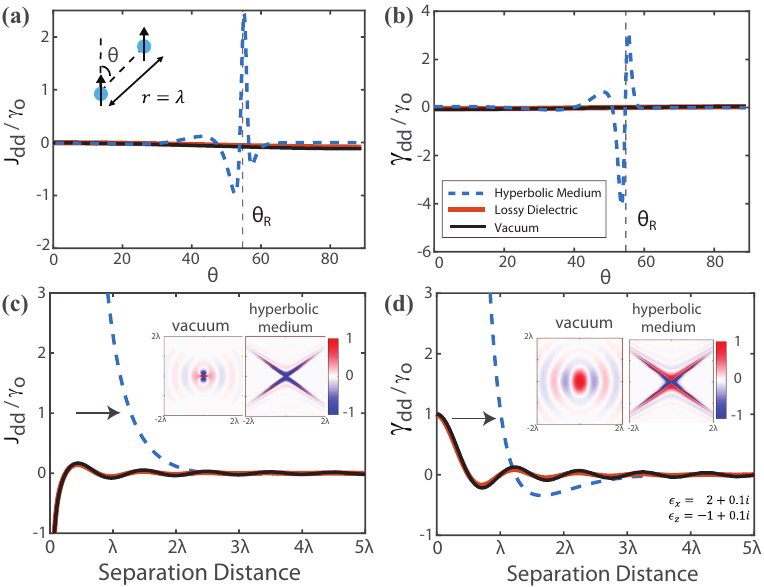}
\end{center}
\vspace{-10px}
\caption{Angular dependence of (a) cooperative Lamb shift (CLS) $J_{dd}$ and (b) cooperative decay rate (CDR) $\gamma_{dd}$ for two z-oriented dipoles in a lossy hyperbolic medium, lossy dielectric, and vacuum. The CLS and CDR have large peaks near the resonance angle of the hyperbolic medium indicative of the super-Coulombic interaction, even for distances of a wavelength.  Comparison of (c) CLS and (d) CDR at the resonance angle versus interatomic separation distance. The CLS and CDR both obey a $1/r^3$ power law dependence in the near-field due to the inclusion of absorption in the hyperbolic medium. Note that the giant interactions start occuring at distances on the order of a wavelength  (arrows) even in the presence of material absorption which is in stark contrast to vacuum. The insets show the contrasting spatially-resolved (c) CLS and (d) CDR for vacuum and for a hyperbolic medium.}
\label{figure2} 
\end{figure}
Figure (2) shows the result of the cooperative Lamb shift and decay rate for two $z$-oriented dipoles in a hyperbolic medium that includes material absorption. We compare the resonant dipole-dipole interactions with the conventional results of a lossy dielectric and vacuum. Note that the RDDI peaks near the resonance angle $\theta_R$ as predicted theoretically. The spatial field plots in the insets clearly demonstrate the distinguishing features of the RDDI in a hyperbolic medium compared to vacuum. Fig (2c) and (2d) demonstrate the $r_e^{-3}$ Super-Coulombic spatial dependence along the resonance angle. Note that the sign of the interaction is dependent on the orientation of the dipoles as well as the relative position of the dipoles within the hyperbolic medium. 

We now turn to the unique orientational dependence of the RDDI between two atoms positioned along the resonance angle $\theta_R$. In figure (3), we plot the normalized cooperative Lamb shift of two atoms a full wavelength apart ($r=\lambda$) as a function of dipole orientation angle $\phi$. The cooperative Lamb shift has a minimum when $\phi=\theta_R$ and a maximum when $\phi=\theta_R+\pi/2$. Assuming that $|\epsilon'|=|\epsilon'_{x}|\approx|\epsilon'_{z}|$, $\epsilon''=\epsilon''_{x}\approx\epsilon''_{z}$, and $\epsilon''\ll|\epsilon'|$, we find that the ratio between the maximum and minimum is $\frac{J_{dd}(\phi=\theta_R+\pi/2)}{J_{dd}(\phi=\theta_R)} \approx  -\frac{3}{2}\left(\frac{\epsilon'}{\epsilon''}\right)^2,$
showing that it is proportional to the square of the figure of merit  of the hyperbolic medium. In figure (3), we use the full Green's function to calculate the orientational dependence of the dipolar interaction in a hyperbolic medium with material absorption, and find excellent agreement with the analytical expression.\\ 
\begin{figure}[t!]
\begin{center}
\includegraphics[width=70mm]{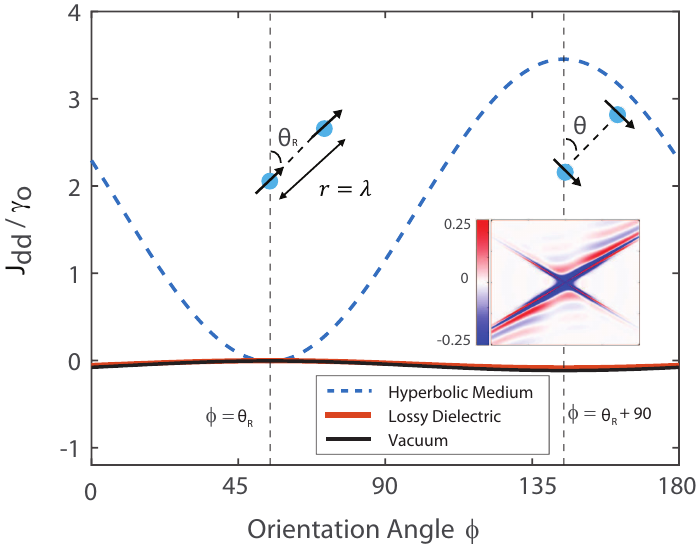}
\end{center}
\vspace{-15px}
\caption{Unique orientational dependence of RDDI in hyperbolic media. The plot shows CLS versus orientation angle $\phi$ for two dipoles positioned along the resonance angle. The cooperative Lamb shift is minimized when the dipoles are collinear with the the resonance angle, and it is maximized when the dipoles are perpendicular to the resonance angle. The inset shows the asymmetric nature of the spatially-resolved $J_{dd}/\gamma_o$ when the dipoles are orthogonal to the resonance angle.}
\label{figure3} 
\end{figure}
We now consider 2$^{nd}$ order super-coulombic QED interactions arising from initial state preparation consisting of atom A in its excited state and atom B in its ground state, $\ket{i}=\ket{e_A}\ket{g_B}$. In the weak-coupling regime, an irreversible resonance energy transfer takes place transferring a photon from atom A to atom B. This process is F\"orster resonance energy transfer (FRET) and the transfer rate given by Fermi's golden rule is $\Gamma_{_{ET}}=2\pi\hbar^{-1}|V_{dd}|^2 \delta(\hbar\omega_A-\hbar\omega_B)$. Along the resonance angle, FRET is mediated by hyperbolic modes and the rate is given by
\begin{equation}
	\Gamma_{_{ET}} \approx \frac{2\pi}{\hbar}\frac{|\mathbf{d}_B\cdot \boldsymbol\kappa_{nf}\cdot\mathbf{d}_A|^2}{(4\pi\epsilon_o)^2|\epsilon_{x}||r_e|^6}\delta(\hbar\omega_A-\hbar\omega_B)
	\label{FRET}
\end{equation}
which shows a $r_e^{-6}$ scaling dependence -- the key signature of second order Super-Coulombic interactions in hyperbolic media.  In addition to the FRET rate, there is also a predicted frequency shift that comes from the initial state preparation $\ket{i}=\ket{e_A}\ket{g_B}$. This is the excited-state Casimir-Polder potential, $U_{eg}(r) = U_{eg}^{r}(r) + U_{eg}^{or}(r)$, composed of a resonant and off-resonant contribution. The resonant excited-state Casimir-Polder potential is of the form $U_{eg}^r(r) = -\frac{|\mathbf{d}_A|^2\omega_A^4}{3\epsilon_o^2 c^4} \alpha_B(\omega_A)\text{Re}\{\text{Tr}[\mathbf{G}(\mathbf{r}_B,\mathbf{r}_A;\omega_A)\mathbf{G}(\mathbf{r}_A,\mathbf{r}_B;\omega_A)]\}$ \cite{safari2015body}.  We therefore predict that the excited-state energy potential will also diverge with a $r_e^{-6}$ scaling dependence similar to the FRET rate. 
Figure (4) shows the full numerical results for the 2nd order dipole-dipole interactions in a lossy hyperbolic medium, a lossy dielectric, and vacuum. In the non-retarded regime ($r\ll\lambda$), we clearly see the effect of the Super-Coulombic interaction which results in a large enhancement of the dipolar interactions $U_{eg}$ and $\Gamma_{ET}$ (shown in inset).  

\begin{figure}[t!]
\begin{center}
\includegraphics[width=70mm]{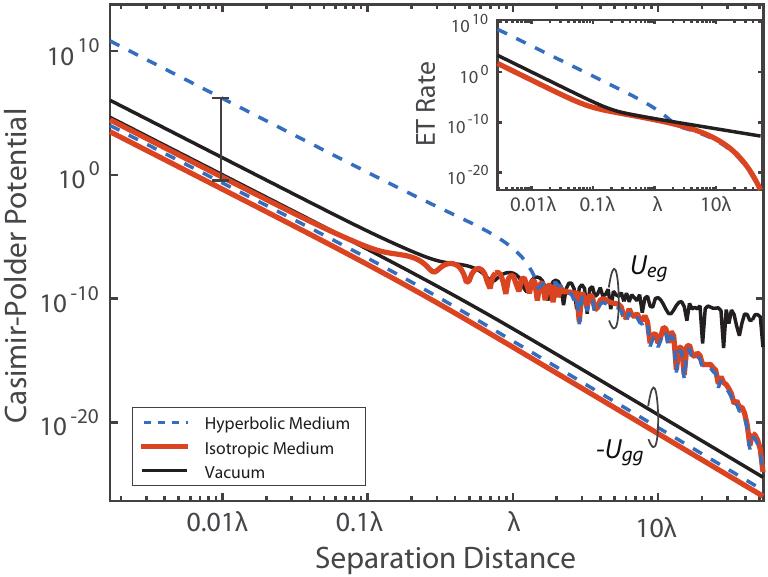}
\end{center}
\vspace{-15px}
\caption{Casimir-Polder interaction energy between two ground-state atoms ($U_{gg}$) and between an excited-state atom and ground-state atom ($U_{eg}$) show fundamental differences when interacting in hyperbolic medium.  $U_{eg}$ $>>$  $U_{gg}$ since resonant interactions lie completely within the bandwidth of hyperbolic dispersion and are strongly enhanced.  The results are normalized to $U_{gg}$ in vacuum,  evaluated  at the near-field interatomic distance of $r_o=\lambda/100$. The inset shows the giant enhancement of the FRET rate, $\Gamma_{ET}$, as compared to vacuum. The FRET rate is normalized to the vacuum energy transfer rate evaluated at $r_o$. }
\label{figure4} 
\end{figure}

It is interesting that the dispersive Van der Waals interaction between two ground state atoms does not diverge in a hyperbolic medium. Using fourth order perturbation theory \cite{safari2006body}, the interaction energy between two ground-state atoms is given by $U_{gg}(\mathbf{r}_A,\mathbf{r}_B) = \frac{-\hbar\mu_o^2}{2\pi}\!\int_0^\infty\!\!d\eta \eta^4\alpha_A(i\eta)\alpha_B(i\eta) \text{Tr}[ \mathbf{G}(\mathbf{r}_B,\mathbf{r}_A;i\eta)\mathbf{G}(\mathbf{r}_A,\mathbf{r}_B;i\eta)]$, where $\alpha_{A,B}(\omega)$ is the isotropic electric polarizability of atom A or B. In the non-retarded limit, the dominant contribution is given by
\begin{equation}
	U_{gg}\approx-\frac{\hbar}{32\pi^3\epsilon_o^2 }\int_0^\infty\!\!\!d\eta\, \frac{\text{Tr}[\boldsymbol{\kappa}_{nf}^2(i\eta)]}{\epsilon_{x}(i\eta)r_e^6(i\eta)}\alpha_A(i\eta)\alpha_B(i\eta)
	\label{VdW}
\end{equation}
which reduces to the well known free-space non-retarded Van der Waals interaction energy when $\epsilon_{x}=\epsilon_{z}=1$. It is important to note that the integral is performed over the entire range of positive imaginary frequencies ($\eta=i\omega$). Generally, the hyperbolic condition $\epsilon_{x}\epsilon_{z}<0$ is only satisfied within a finite bandwidth of the electromagnetic spectrum. We therefore expect that it would not alter the broadband cumulative effect of the entire electromagnetic spectrum, and as a result we predict that the ground-state ground-state interaction energy will not diverge in a hyperbolic medium. From fig. 4, it is also clear that the ground-state ground-state Casimir Polder potential $U_{gg}$ does not show any type of enhancement for the hyperbolic medium, in agreement with our discussion. Note that the scaling dependence in the non-retarded regions is in agreement with equations (\ref{FRET})-(\ref{VdW}), as expected. In the retarded regime ($r\gg\lambda$), the excited-state interactions $U_{eg}$ and $\Gamma_{_{ET}}$ display an exponential damping behaviour due to material absorption, while the ground-state interaction $U_{gg}$ displays the typical Casimir-Polder power law dependence, $r^{-7} $ (Fig. 4).

\begin{figure}[t!]
\vspace{-10px}
\begin{center}
\includegraphics[width=85mm]{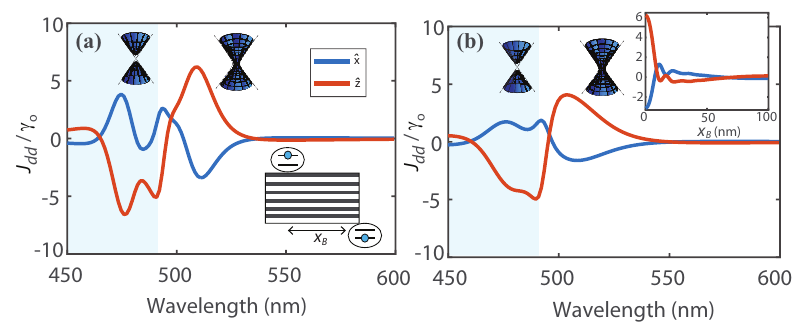}
\end{center}
\vspace{-20px}
\caption{ Giant long-range Cooperative Lamb shift, $J_{dd}$, across a practical plasmonic super-lattice calculated for the corresponding (a) effective hyperbolic model and (b) a 40-layer multilayer system taking into account dissipation, dispersion and finite unit cell size. Atom A is 4 nm away from top interface, while atom B is adsorbed to bottom interface with a horizontal displacement of $x_B= 5$ nm. The total slab thickness is 100 nm. The inset shows the cooperative Lamb shift dependence on atom B's horizontal displacement. The red and blue curves denote the two orientations of the transition dipole moment of the atoms. Good agreement is seen between the EMT model and practical multilayer design paving the way for an experimental demonstration of the super-Coulombic effect with cold atoms.}
\label{figure5} 
\vspace{-5px}
\end{figure}
Fig. (5) proposes a practical plasmonic super-lattice system to enhance atom-atom interactions taking into account the role of dissipation, dispersion and finite unit cell size. We show the large enhancement of cooperative Lamb shift $J_{dd}$ for an effective medium model and compare it to a 40-layer structure consisting of $Ag$ and $TiO_{2}$ with a total slab thickness of $100$ nm. The two large peaks seen in fig. (5) occur when the dispersive resonance angle $\theta_R(\lambda)$ is equal to the fixed separation angle, i.e. $\theta_R(\lambda) = \theta_o$ in agreement with theory. 


To conclude, we revealed a class of singular excited-state atom-atom interactions in hyperbolic media that arises from a fundamental modification of the coulombic near-field. The experimental observation of such effects will require careful isolation of the medium induced cooperative interactions between atoms from the effect of single atoms interacting with the hyperbolic medium. Our work paves the way for studies of long-range entanglement and self-organization \cite{gonzalez2015subwavelength} and is also a first step towards cold atom studies with hyperbolic media exhibiting unique effects that are not found in photonic crystals or cavities. 

\vspace{-20px}

\bibliography{SuperCoulombic11}

\end{document}